# Multilayer Perceptron and Geometric Albedo Spectra for Quick Parameter Estimations of Exoplanets

**Timothy K Johnsen[1,2] and Mark S Marley[2]**

[1] SETI Institute (189 N Bernardo Ave suite 200, Mountain View, CA 94043)
[2] NASA Ames Research Center (Moffett Blvd, Mountain View, CA 94035)

E-mail: Tim.K.Johnsen@gmail.com Mark.S.Marley@NASA.gov

**Abstract**

Future space telescopes now in the concept and design stage aim to observe reflected light spectra of extrasolar planets. Assessing whether given notional mission and instrument design parameters will provide data suitable for constraining quantities of interest typically requires time consuming retrieval studies in which tens to hundreds of thousands of models are compared to data with a given assumed signal to noise ratio, thereby limiting the rapidity of design iterations. Here we present a novel machine learning approach employing five Multilayer Perceptron's (MLP) trained on model albedo spectra of extrasolar giant planets to estimate a planet's atmospheric metallicity, gravity, effective temperature, and cloud properties given simulated observed spectra. The stand-alone C++ code we have developed can train new MLP's on new training sets, within minutes to hours, depending upon the dimensions of input spectra and desired output. After the MLP is trained, it can classify new input spectra within a second, potentially helping speed observation and mission design planning. Four of the MLP's were tuned to work with various levels of spectral noise. The fifth MLP was developed for cases where the user is uncertain about the noise level. The MLP's were trained using a grid of model spectra that varied in metallicity, gravity, temperature, and cloud properties. We tested the MLP on noisy models and observed spectra of both Jupiter and Saturn. The root mean squared error when applied to noisy models were on the order of the model grid intervals. The results show that a trained MLP is a robust means for reliable in-situ estimations, providing an elegant artificial intelligence that is simple to customize and quick to use.

Keywords: MLP, Geometric Albedo Spectra, Exoplanets, Machine Learning

## 1: Introduction

Exoplanet spectroscopic science to date has focused primarily on transiting planets (Kreidberg 2018) and a handful of young giant planets directly imaged in thermal emission (e.g., Nielsen et al. 2019). Next generation space telescopes, including the coronagraph instrument on WFIRST (Mennesson et al. 2018) and mission concepts such as the Large Ultraviolet Optical InfraRed Surveyor (LUVOIR) and the Habitable Planets Explorer (HabEx) space telescopes, will have the capability of obtaining photometry and spectroscopy of directly imaged planets in reflected light. Design of mission concepts and notional observing plans for such telescopes requires establishing the desired signal-to-noise ratio and spectral resolution required to meet specified science goals, such as determining the atmospheric metallicity of an extrasolar giant planet. To date such requirements have been set by iterative retrieval studies of simulated spectra (e.g. Brandt 2014; Lupu et al. 2017; Nayak et al. 2018; Feng et al. 2019) or information content studies (Batalha et al. 2018). Retrieval studies are very time consuming as they involve tens to hundreds of thousands of iterative model spectra to be evaluated in order to ascertain which ranges of models best fit the simulated datasets. Retrieval approaches are thus ill-suited for applications involving iterative design studies which aim to specify instrument capabilities or observation parameters. For example, a user may desire to understand if a given observation of a target with a given integration time and other instrument settings, simulated with an online observation simulation tool[1], would be sufficient to answer a particular science question. Running a full retrieval of the simulated data in such a situation is generally not practical. Here we report the development of a machine learning tool to facilitate rapid data interpretation given a simulated observation. Our goal is to understand if such a tool could be used for quick estimations of planetary properties, not to replace retrievals.

Machine learning algorithms have previously been applied to exoplanetary spectra such as these. Waldmann (2016) used stacked Restricted Boltzmann Machines topped with a perceptron, called RobERt, to identify gasses in simulated highly irradiated exoplanet emission spectra. Zingales and Waldmann (2018) also aimed to address what they termed the "computational bottleneck" associated with retrieval methods by developing a Generative Adversarial Network, called ExoGAN, to estimate mass, radius, temperature and chemical abundances from hot Jupiter transit data. Màrquez-Neila (2018) used random trees to estimate temperature and chemical abundances from simulated transit datasets. Soboczenski et al. (2018) used a convolutional network to estimate chemical abundances of simulated spectra of terrestrial exoplanets in reflected light comparable to those expected from LUVOIR. Cobb et al. (2019) used an ensemble of Bayesian neural networks, called plan-net, to accurately predict the isothermal temperature and water abundance using a transmission spectrum of WASP-12b. Notably absent from these previous studies are applications to a set of extrasolar giant planets in reflected light, such as we consider here.

Specifically, we begin with a training set drawn from approximately 50,000 model geometric albedo spectra of cool extrasolar giant planets computed by MacDonald et al. (2018) and available online[2]. The models are down sampled in resolution to be comparable to that expected to be obtained by future large space observatories which aim to measure reflected light spectra of terrestrial and giant planets.

We present a Multilayer Perceptron (MLP) to add to the array of machine learning algorithms used to classify exoplanets. The MLP we present has fewer trainable parameters then previous approaches, reducing both complexity and computation time. As in previous studies the models are distinguished by effective temperature, gravity and metallicity. Unlike the previous studies which considered cloudless spectra we also train on the cloud parameter $f_{sed}$ (Ackerman & Marley 2001), a measure of cloud vertical thickness. Small values are associated with decreased sedimentation efficiency and vertically thicker and more optically thick clouds. Conversely larger values are associated with thinner clouds and spectra which approach the cloudless limit.

The methods presented in this paper were adapted from similar methods that were previously used to classify minerals with Raman spectra and machine learning (Ishikawa and Gulick 2013; Johnsen and Gulick 2019).

**2: Philosophy**

The purpose of this study was to provide a quick and easily accessible and understandable tool for scientists in the preliminary stage of investigating exoplanet reflectance spectra. Thus, the machine learning algorithm we investigated in this study was an MLP selected to have a relatively low complexity, high portability, and quick train/test times while maintaining an accuracy appropriate for the job of order-of-magnitude on-the-fly estimations. Current approaches (Waldmann 2016; Waldmann 2018; Màrquez-Neila 2018; Soboczenski et al. 2018; Cobb et al. 2019) were out of scope of this study as they may have higher accuracies at the cost of either increased complexity or run time; or the approaches used either different input/output parameters or input sources.

We chose a neural network as our machine learning family because it can output a continuous value for each parameter, allowing to interpolate in between the grid-space of trained parameters. A neural network paired with deviated error, Root Mean Squared Error (RMSE), can quickly estimate a continuous parameter given an input spectrum, with estimated error bounds. Other algorithms, such as random trees, output posterior distributions of predicting parameter values that fall into the discrete parameter values trained on. The advantage of using a neural network is one network can output one continuous value with given error bounds for each parameter, as opposed to random trees which use multiple trees to output a posterior distribution for each parameter. A combination of the two may be an ideal machine learning ensemble.

We chose to use an MLP, in the neural network family, because it has fewer variables which need to be trained, such as weights and bias terms, as compared to more complex approaches in the neural network family. This drastically reduces time needed to train a neural network, allowing quick optimizations and predictions based on input dimensions of the spectra and desired output parameters, which both may vary by the scientists' preliminary investigations and goals. Future studies will need to consider utility of other approaches.

---

[1] e.g., http://luvoir.stsci.edu/coron_model

[2] https://zenodo.org/record/1210305#.XUjTlJNKjUI

## 3: Database

As a database we chose the geometric albedo spectra ($A_g$) of cool gas giant extrasolar planet models computed by MacDonald et al. (2018). Such planets will make bright, appealing targets for future direct imaging observations and their distinctive optical spectra are well suited to test machine learning algorithms. These reflectivity spectra for planets observed at full phase were computed employing the modelling approach of Marley & McKay (1999) as extended to exoplanet spectroscopy (Marley et al. 1999; Cahoy et al. 2008). In reality exoplanet reflected light spectra will neither be available at full phase (see Nayak et al. 2018 for retrievals against such data) nor cast as albedo spectra since the planetary radius will not be known. Nevertheless, we make both approximations to make the immediate problem more tractable and the capabilities of the MLP to be more straightforwardly evaluated. Each model is for a particular value of atmospheric metallicity [M/H], gravity, log $g$(cm s$^{-2}$), effective temperature, $T_{eff}$, and cloud sedimentation efficiency $f_{sed}$ (Ackerman & Marley 2001).

The original model set by MacDonald et al. consists of about 50,000 individual geometric albedo spectra. Here we limited [M/H] from 0.0 to 2.0 with intervals of 0.5, log $g$(cm s$^{-2}$) from 2 to 4 with intervals of 0.1, $T_{eff}$ from 150 to 400 K with intervals of 10 K, and $f_{sed}$ from 1 to 10 with intervals of 1. We choose this range to better capture the planets most easily detected by direct imaging missions as the warmer planets will both likely be too close to their stars to be detectable and are generally dark in reflected light as they lack clouds (Marley et al. 1999; Sudarksy et al. 2000; Angerhausen et al. 2015; Heng & Demory 2013). When cloud opacity is negligible, $f_{sed}$ does not appreciably affect the spectra and there are thus 10 model degeneracies for a given [M/H], log $g$(cm s$^{-2}$), and $T_{eff}$. For this reason, the models with negligible cloud levels were discarded. Because of the model degeneracies, an ideal machine learning algorithm would compute posterior distributions, however a single MLP is incapable of this and we did not explore further. This resulted in 12,820 models in our entire database.

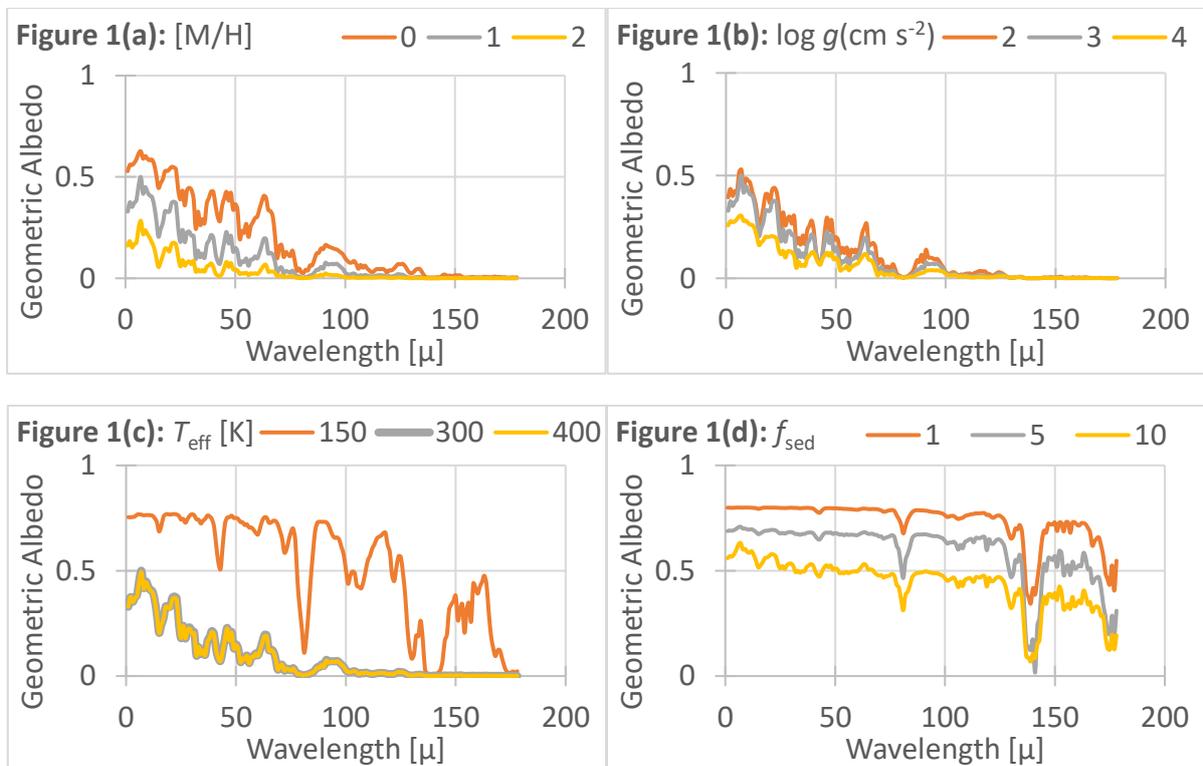

**Figure 1**: Model albedo spectra of generated exoplanets. Each panel shows a range of model spectra over the domain considered here for that given parameter. The grey spectrum is the same for panels (a) through (c): [M/H] = 1.0, log(g) = 3.0 cm s$^{-2}$, $T_{eff}$ = 300 K, and $f_{sed}$ = 5. The orange and yellow spectra in each figure are respectively the smallest and largest values for the given output parameter, assuming the other parameters remain the same as the grey spectrum. However, panel (d) uses [M/H] = 1.0, log(g) = 3.5 cm s$^{-2}$, and $T_{eff}$ = 190 K to better highlight differences. There are still some degeneracies in the model spectra, for example panel (c).

*3.2: Adding Gaussian Noise to Model Spectra*

We simulated uncorrelated noise in the model spectra at various levels. Random, Gaussian noise was added by individually calculating the average albedo of a spectrum and using a percentage of that average as the standard deviation, σ, for the random distribution. We synthesized noise for every data point by adding a number normally distributed with the corresponding σ for a given spectrum. Albedo values were bound between 0 and 1. If random noise would otherwise exceed this threshold, the values were set to either a 0 or 1, whichever was closer. We used three levels of noise: setting σ equal to either 5%, 10%, or 20% of the average, corresponding to signal to noise ratios of 20, 10, and 5, respectively. Figure 2 shows an example of Gaussian noise added to a model spectrum.

Noise draws were randomly computed ten times for each model spectrum at each noise level. An original model spectrum was thus associated with 30 additional noisy models (10 for each noise level). Note that we did not attempt to emulate any noise model for a particular notional instrument as these frequently change. However, such a modification could easily be included in our approach. Noisy spectra were not used to train an MLP. Instead, the MLP's were trained on the models without noise, and tested on different levels of noisy spectra.

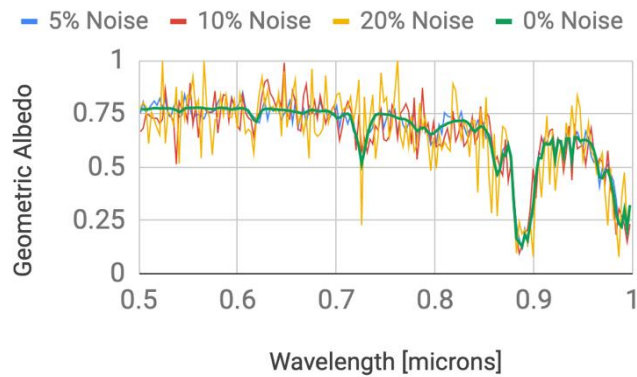

**Figure 2**: Examples of model spectra with various levels of noise added. The Gaussian noise is random, uncorrelated in wavelength, and added to a model spectrum ([M/H] = 0.0, log(g) = 2.0 cm s$^{-2}$, $T_{eff}$ = 150K, and $f_{sed}$ = 1). Noise particularly impacts the weaker CH4 spectral features in this example.

*3.3: Albedo Spectra of Jupiter and Saturn*

We also included geometric albedo spectra of Jupiter and Saturn (Karkoschka et al. 1999) in our test set. The Jupiter and Saturn spectra, after being binned to the same spectral resolution as the models, are shown in Figure 3.

The effective temperature of Jupiter and Saturn (~130 and 95K respectively) are cooler than the range of exoplanet models. These are included in our test to understand how the MLP handles such cases beyond the edge of the available model set where $T_{eff}$ has a large influence on the shape of the model spectrum (see Figure 1). However, Jupiter and Saturn both lie within the domain for other model parameters.

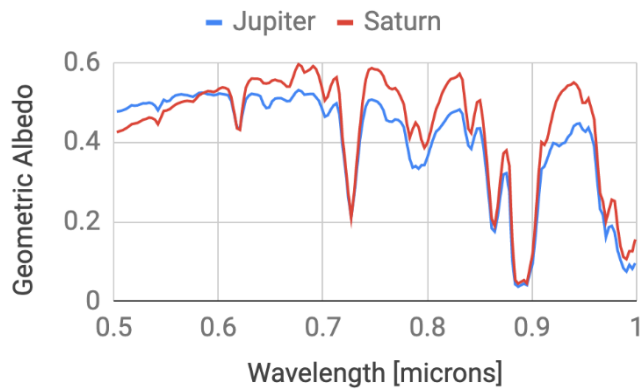

**Figure 3**: Geometric albedo spectra of Jupiter and Saturn from Karkoschak et al. (1999).

*3.4: Spectral Wavelength Range and Resolution*

As the number of datapoints input into the MLP increases, the complexity of the problem exponentially increases as each new input dimension requires a new set of weights to connect to the rest of the neural network's nodes. This phenomenon is commonly referred to as the curse of dimensionality (Bellman 1966). We took two steps to decrease the number of datapoints to decrease the training time while maintaining significant features in the spectra.

We first used a smaller wavelength, $\lambda$, range in the spectra than available, $0.5016 \leq \lambda < 1.0$ μm (wavelengths of 0.50 and 1.00μm were not included because of an occasional glitch in the tabulated models at these wavelengths). To further reduce the dimensionality of the spectra, we averaged consecutive intensities into bins to decrease the spectral resolution. To preserve the spectral information while making the MLP more tractable we chose 178 bins, corresponding to a spectral resolution R = $\lambda/\Delta\lambda$ ~270 at 0.75 μm. Bins are referred to by the midpoint of the wavelength range for that bin. Other decomposition and machine learning techniques can be used to reduce dimensionality, however we found binning to be sufficient.

*3.5: Matrix Notation*

Let S be an N x I matrix of spectra for a given dataset. Each row in S ($s_{n,*}$) is a different spectrum where each element ($s_{n,i}$) is the albedo value for the i$^{th}$ component. Each S has a corresponding N x K matrix, L, containing labels for the spectra. Each row in L ($l_{n,*}$) is the respective label for the appropriate spectrum ($s_{n,*}$), where each element ($l_{n,k}$) is a parameter value for that given spectrum: [M/H], log $g$(cm s$^{-2}$), $T_{eff}$, and $f_{sed}$. A pair of corresponding matrices for an example dataset is shown in Table 1. Each element in S represents an average albedo value over a binned wavelength range. The albedo matrix corresponds to a matrix of labels, L, and an example is in Table 2.

**Table 1**: Example spectral matrix.

| $A_g$ (0.503 μm) | ... | $A_g$ (0.998 μm) |
|---|---|---|
| $s_{1,1}$=0.772 | ... | $s_{1,I}$=0.322 |
| ... | ... | ... |
| $s_{N,1}$=0.107 | ... | $s_{N,I}$=0.811 |

**Table 2**: Example labels matrix.

| [M/H] | Log $g$(cm s$^{-2}$) | $T_{eff}$ [K] | $f_{sed}$ |
|---|---|---|---|
| $l_{1,1}$=0.0 | $l_{1,2}$ = 2.0 | $l_{1,3}$ = 150 | $l_{1,4}$ = 1 |
| ... | ... | ... | ... |
| $l_{N,1}$=2.5 | $l_{N,2}$ = 4.0 | $l_{N,3}$ = 400 | $l_{N,4}$ = 10 |

**4: Multi-Layer Perceptron**

A neural network is a machine learning algorithm that uses a network of connected, weighted nodes to fit input to output. The network is supervised, in the sense that it trains on rows in **S** to adjust the weights to better fit to **L**. With trained weights, the network can estimate parameters of novel spectra.

Figure 4(a) illustrates the type of neural network we used, a Multi-Layer Perceptron (MLP) (Bishop 1995; Ripley 1996; Haykin 1998; Fine 1999). Each color indicates a different layer or nodes. The red, input, layer is used to input a row of **S**. Each input node reads a different element in the row. The set of orange, hidden, layers is the heart of an MLP. Each hidden node is an obfuscated feature that the MLP must learn by summing weighted input nodes and inputting the sum into an activation function. There can be multiple hidden layers with a different number of nodes in each layer, referenced by the set H. The output, green, layer consists of output parameters we are trying to map from input. Output nodes work like hidden nodes, by summing weighted hidden nodes from the previous layer and inputting the sum into an activation function. Different activation functions can be used depending on the purpose and limitations of data flowing into and out of that node.

Figure 4(b) shows the activation function that we used for hidden nodes, Exponential Linear Units (ELU) (Clevert et al. 2016). ELU functions are simple linear functions for positive weighted sums, output is bound between -1 and positive infinity. ELU is an updated version of Linear Units (LU) which is a linear function bound between 0 and positive infinity. The training process, as detailed later, relies on derivatives to update weights. Giving ELU a simple diminishing output for negative sums improves the training process over LU, because training can better escape negative values.

Figure 4(c) shows the activation function that we used for output nodes, Sigmoid, otherwise known as the logistic function. Sigmoid activation functions were used for the output nodes to bind output values to the range used for each parameter. For example, we trained the network on spectra modeled from planets with varying temperature between 150 and 400 K, we can stretch the sigmoid function so that it asymptotically reaches 140 K as a lower bound (giving a buffer of 10 K as determined by the intervals between 150 to 400 K) and similarly an upper bound of 410 K.

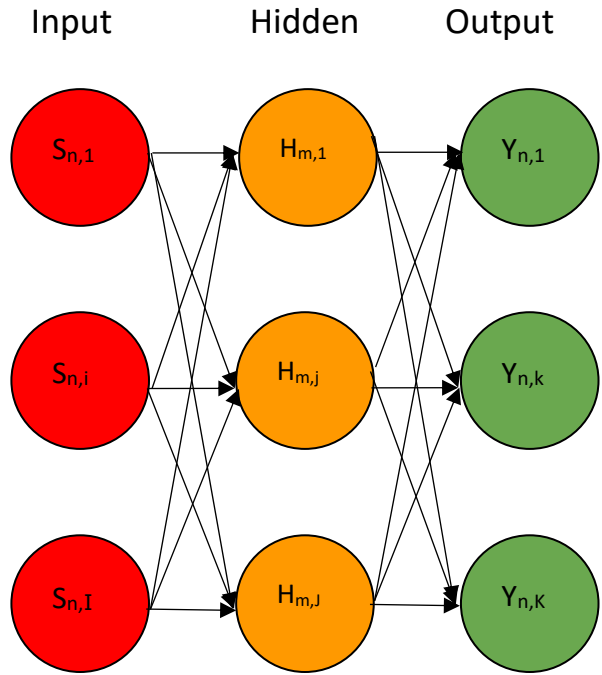

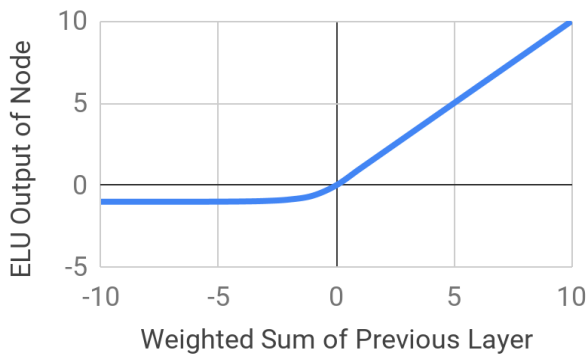
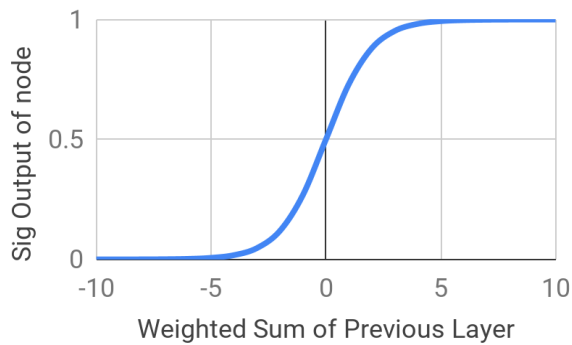

**Figure 4(a):** MLP. An input row $s_{n,*}$ feeds data to hidden nodes $\mathbf{H} = \{h_{m,1} \ldots h_{m,J\_m}\}$ to predict an output row $y_{n,*}$.
**Figure 4(b):** ELU activation function, $y = x$ for positive input values and $y = e^x - 1$ for negative input values.
**Figure 4(c):** Illustration of a Sigmoid activation function, $y = 1/(1+e^{-x})$.

Let M be the total number of layers including input, hidden and output. Quantity m varies from 1 to M, representing the indexed layer in the network where the input layer corresponds to m=1 and output layer corresponds to m=M. The hidden layers have the indices m=2 to m=(M-1).

The number of nodes in each layer can vary. The input layer has I number of nodes, as determined by the dimension I of S. The number of nodes in each hidden layer must be determined by the user, as presented in section 5. We used up to 4 output nodes, one for each label: [M/H], log $g$(cm s$^{-2}$), $T_{\text{eff}}$, and $f_{\text{sed}}$.

There are trainable parameters that the MLP must learn to map input to output. These parameters consist of a set of weights, **W**, and bias terms, **B**, with elements in the sets indexed to each node by the layer number, m, and node number in that layer, j (such as $w_{m,j}$ and $b_{m,j}$). Quantity $J_m$ refers to the number of nodes in that layer. Weights are used to take a weighted sum of nodes in the previous layer. Bias terms are used as a shift in the activation function. With weights and bias terms, data propagates forward from input layer to hidden layer using equation 1, hidden to another hidden using equation 2, and hidden to output with equation 3. Where f is the activation function. The flow of data, as propagated forward with equations 1-3, is illustrated in Figure 4(a).

$$h_{2,j} = f\left(b_{2,j} + \sum_{i=1}^{J_1 = I}(w_{1,i} * s_{n,i})\right) \quad (1)$$

$$h_{m,j} = f\left(b_{m,j} + \sum_{i=1}^{J_{m-1}}(w_{m-1,i} * h_{m-1,i})\right) \quad (2)$$

$$y_{n,k} = f\left(b_{M,k} + \sum_{i=1}^{J_{M-1}}(w_{M-1,i} * h_{M-1,i})\right) \quad (3)$$

The grid of trainable MLP parameters must be explored to find the optimal values that give lowest output error. Supervised machine learning algorithms iteratively introduce training vectors, with known labels, to the MLP. The MLP trains on the feature vectors by exploring the grid space to find values that minimize the error between output values and labels. Stochastic Gradient Descent (SGD) (Robins et al. 1951) describes the process used to explore the grid space, as detailed below.

Trainable MLP parameters in equations 1 and 2, **Θ** = {**W**, **B**}, were at first randomized to small values normally distributed around 0 with a standard deviation, STD, found from equation 3. This gives a near zero deviation which scales to the size of the MLP. Near zero values were adequate starting points for the grid search, as zero is a critical point in both ELU and Sigmoid activation functions.

$$STD = 1/\sqrt{I + K} \quad (4)$$

We then iteratively introduced a subset of training spectra to the MLP. After each iteration, a loss function was used to measure the error between labels and output values. We used mean squared error as our loss function, equation 5. The 0.5 coefficient was used so that it vanishes after taking the derivative. Squared error was used because: 1) it measured error in continuous values as opposed to other loss functions which focus on binary classification error; and 2) it punishes larger residuals, as opposed to other loss functions which balance out small and large residuals more.

$$\varepsilon_k = 0.5\,(l_{n,k} - y_{n,k})^2 \quad (5)$$

After each iteration the average error, $\varepsilon$, for each output node was calculated for 128 training spectra. After measuring error each iteration, **Θ** was updated to lower error. The gradient of $\varepsilon$ with respect to each parameter, $\frac{\partial \varepsilon}{\partial \theta}$, was averaged over the 128 spectra. By averaging the gradient with these "mini batches" of 128 spectra each iteration, we were able to parallelize the training process which sped up training (Li 2014).

The quantity $\frac{\partial \varepsilon}{\partial \theta}$ was used to update **Θ** after each iteration. Each update adjusted **Θ** in the gradient's direction scaled by the learning rate, η, and adjusted by the learning momentum, α. If η is too large, training can overshoot a global minimum in error. If η is too small, training can get stuck at a local minimum in error. Quantity α was used to help get out of local minima in error by adding a percent of the previous update.

Regularization techniques were also employed to keep the values of **Θ** close to zero. Otherwise, the values could blow up, causing the MLP to overfit on training data. This is analogous to how a polynomial fit with high coefficients gives erroneous results on novel data. $L_1$ and $L_2$ regularization constants were used for this purpose.

The update size for each parameter, $\Delta\theta$, was determined by equation 6. The ± in front of $L_1$ and $L_2$ indicates which ever pulls the MLP parameter closer to zero.

$$\Delta\theta = \alpha * \Delta\theta_{last} - \eta * <\frac{\partial \varepsilon}{\partial \theta}> \pm L_1 \pm L_2 * \theta \quad (6)$$

We utilized a few popular techniques that further improved our grid search. The following terminology refers to an epoch which means iterating through the entire training set one time.

Smith (2015) presents a technique to cycle between different values of η, called cyclic learning. First, one finds a small range of optimal learning rates. The grid search cycles between the minimum and maximum optimal learning rates. Cyclic learning alleviates the need to over-optimize the learning rate, and helps training get over saddle points in the loss function quicker. To cycle through learning rates, we used a linear triangle function with a cycle step size of 2 times the number of iterations in each epoch, as shown by Smith to be robust.

Srivastava et al. (2014) used a pseudo ensemble of neural networks that essentially took the weighted average of nodes, called dropout. During training, input and hidden nodes are initially turned off and only turned on with a given probability each epoch. After training, when using the MLP to make predictions, each node is weighted by the probability it was turned on. This is a quick and synthetic way of averaging the response of multiple trained networks. It can also be used as a regularization technique, as parameters are not trained as often and are scaled down by their respective probability of being turned on during training.

Srivastava et al. (2014) also showed that using another regularization technique, Max-norm, helped in combination with dropout. Max-norm adds an upper limit to weights. Weights for each node are scaled to fit inside an n-dimensional sphere, where n is the number of input connections to that node.

Overfitting is a common problem that arises when a supervised machine learning algorithm has extensively trained on one data set, and overfits the MLP parameters to that one data set. This is why regularization techniques are used to keep the MLP parameters close to zero. It is also why we split our database into three sets: training, validation, and testing. A validation set is used to determine when to stop training. A testing set is a set of vectors that have been left out of the entire process and are used to corroborate results at the end.

After the MLP trains on each row in $\mathbf{S}^{train}$, otherwise called an epoch, training error was measured by reinputting all rows of $\mathbf{S}^{train}$ to predict an output matrix $\mathbf{Y}^{train}$. Then validation error was also measured using the set $\mathbf{S}^{validation}$, by predicting another corresponding output matrix $\mathbf{Y}^{validation}$. Training and validation error were individually measured using Root Mean Squared Error (RMSE), which indicates the deviated error.

$$RMSE = \sqrt{\sum_{k=1}^{K}\sum_{n=1}^{N}(y_{n,k} - l_{n,k})^2 / (K * N)} \qquad (7)$$

As training progresses, RMSE should decrease. This shows the MLP is learning, and properly fitting input to output by optimizing $\Theta$. If the training RMSE continues to decrease while validation RMSE either increases or stagnates, the MLP is not properly learning as it is overfitting to the training data. One must define when to stop training, in which we used two methods. The first stopping method was early stopping, which is a method that stops training before validation RMSE increases or begins to stagnate, effectively stopping training before it overfits. Early stopping is indicated in Figure 5 by where validation error has significantly dropped from its initial value, but before it begins to diverge. The second stopping method was greedy stopping, which is a method that allows training to continue for a given number of epochs and simply uses the values of $\Theta$ which resulted in lowest error.

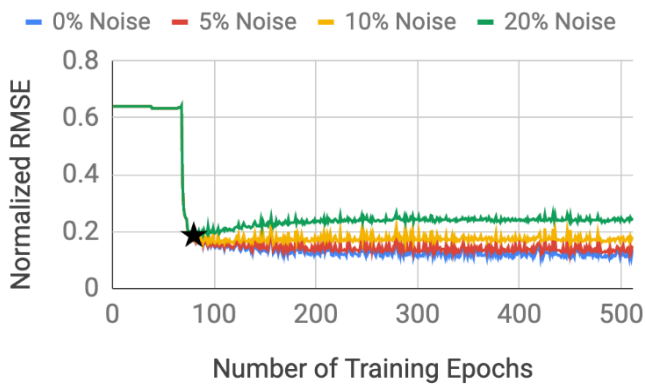

**Figure 5:** Greedy stopping uses the MLP parameters found after several epochs that resulted in the lowest validation RMSE. Early stopping stops training around the black star, where the error begins to diverge, ~80 epochs. At the black star RMSE begins to increase for higher noise levels and fluctuate more for all noise levels, as the number of epochs increases.

The MLP's and all data-processing algorithms were invoked from a lightweight, stand-alone C++ program that we have developed.

**5: Training the MLP on Model Spectra**

Model spectra from our database were randomly split into 3 sets: 80% into a training set, 10% into the validation set, and 10% into the testing set. The training set only included the original model spectra, none of the noisy spectra were added to the training set. Both the validation, and testing set, had all of the synthetically noise-added spectra in them. Noise was left out of the training process, to mitigate any effects of the neural network fitting to the pseudo-random nature of the noise added (i.e. if both the training and testing data generate noise using the same process, then that process is something the neural network could theoretically learn from which biases the machine learning algorithm).

When training a neural network, it is important to include a large number of diverse data in the training set. This way the network can predict a wide range of novel data. A learning curve that plots error versus the size of the training set can be used to determine if enough training data was used. At first the network was completely trained on by only one spectrum from $S^{train}$, then another spectrum was iteratively added to the training set and the network was completely trained again from scratch. Figure 6 is such a learning curve that shows we used enough training spectra in $S^{train}$, because the error reaches an asymptote when using the entire training set.

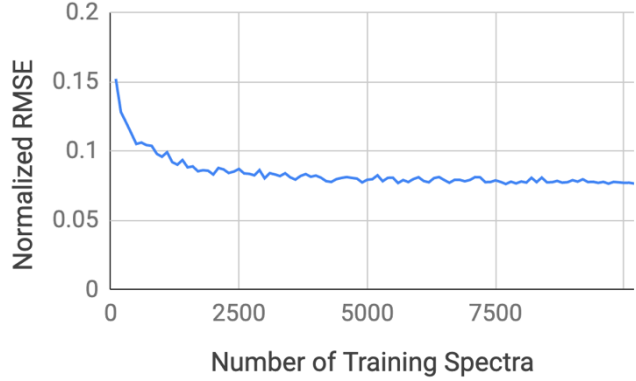

**Figure 6:** Normalized RMSE as a function of the number of models used in the training set. These results indicate that for this model set about four thousand models are required, as we had about 12,000 models in our set then we used an ample number of models.

Labels were normalized between 0 and 1 to fit into the sigmoid function and so that each output parameter ([M/H], $T_{eff}$, g, $f_{sed}$) was similarly scaled when calculating and comparing RMSE. Normalized RMSE is unitless and refers to the total error when using normalized parameters, and absolute RMSE refers to the error in the given units for each parameter.

$$l_{n,k,norm} = [l_{n,k} - \min(l_{*,k})]/[\min(l_{*,k}) - \max(l_{*,k})] \quad (8)$$

$$Normalized\ RMSE = \sum_1^N \sum_1^K \sqrt{\frac{(y_{n,k} - l_{n,k,norm})^2}{K*N}} \quad (9)$$

$$y_{n,k,abs} = y_{n,k}[\min(l_{*,k}) - \max(l_{*,k})] + \min(l_{*,k}) \quad (10)$$

$$(Absolte\ RMSE)_k = \sum_1^N \sqrt{\frac{(y_{n,k,abs} - l_{n,k})^2}{N}} \quad (11)$$

There was a general trend in the training curve for each MLP configuration, see Figure 5. Normalized RMSE would at first stagnate at about 0.6, because MLP parameters were randomly initialized. Then, there would be a sharp drop in RMSE, this indicated when the grid search found an optimal range of MLP parameters. This was followed by a gradual drop in RMSE, sometimes accompanied with a gradual increase in RMSE at higher noise levels. RMSE would then fluctuate around an average showing no improvement.

Early stopping is indicated by the black star in Figure 5. The star represents where the error began to diverge for each of the noise levels. At the star, if the same spectrum was tested with different noise levels, the output values would all be roughly the same. The black star was determined algorithmically by the epoch just before the RMSE began to retrograde for any of the noise levels. We employed early stopping to stop training before the MLP began to overfit on the training data, and to train an MLP that tested the same regardless of the level of noise in the validation spectra (up to 20% noise). If the level of noise is not known this MLP can be applied as it is unbiased to noise.

As an alternative we also utilized greedy stopping which simply lets the MLP train for a number of epochs and took the MLP parameters which resulted in the lowest error. We trained for up to 2048 epochs, as this gave enough training for the validation error from all MLP configurations to converge. Greedy stopping was applied for each of the noise levels. If the user knows the level of noise in the spectra being tested, they can use the optimal MLP for that level of noise.

## 6: Testing the MLP on Noisy Models

We investigated multiple MLP configurations, changing: the number of hidden layers and nodes in each layer, learning rate, learning momentum, and if we used dropout, cyclic learning, L1, L2, and/or Max-Norm. After investigating different MLP configurations, we found that using three hidden layers with 64 nodes in the first hidden layer, 64 in the second, and 32 in the third yielded the lowest validation error. We used a learning momentum of 0.9, as changing it did not significantly affect results. Learning rate cycled between 0.1 and 0.9, as this was the optimal range. For this study we selected the five MLP's listed in Table 3. One MLP from early stopping, and one MLP for each noise level when using greedy stopping, termed as Greedy0-20 for 0-20% noise added to the validation set. Early stopping had all levels of noise in the validation set.

**Table 3:** The five selected MLP's.

| Stopping Approach | # of Epochs | log(L1) | log(L2) | Max-Norm (radius) | Dropout (probability to turn on: input, hidden nodes) |
|---|---|---|---|---|---|
| Early | 295 | -6 | -4 | 6.1 | None |
| Greedy0 | 2010 | -6 | -4 | 6.1 | None |
| Greedy5 | 1814 | -6 | -4 | 1.4 | None |
| Greedy10 | 1791 | -5 | -6 | 1.5 | 1.0, 0.9 |
| Greedy20 | 2026 | -6 | -5 | 1.5 | 0.9, 1.0 |

Validation results of thousands of different MLP's showed a general trend that noisier spectra benefit from stronger regularization techniques which oppositely increases error when using less noisy spectra. This trend is apparent in Table 3 by the large drop in Max-Norm radius when noise levels increase from 0% to 5%. Also, dropout only improves error when using the 10% and 20% greedy MLP's. We assume this is because regularization forces the MLP parameters to be more generalized, which allows more room for uncertainty in the data points.

We used $S^{test}$, with all noisy spectra, to test the MLP's. Figure 7 illustrates the absolute RMSE of each MLP after testing them on $S^{test}$ which had been left out of the entire process up until this point. The absolute RMSE values are also tabulated in Figure 7, and can be directly used as uncertainty values in final output parameters. The MLP's that performed well at lower noise levels, performed poorly at higher noise levels and vice versa.

## 7: Testing the MLP's on Jupiter and Saturn

The five MLP's were also tested on Jupiter and Saturn spectra. Figure 8 demonstrates the predictions for each output parameter. The uncertainty bars were determined by the testing RMSE which resulted from using 0% noise for the respective MLP. The signal to noise level for the Jupiter and Saturn spectra is 1000 (Karkoschka et al. 1999) which is essentially zero noise for our purposes. The minimum and maximum values of the y-axis represent the end points of the model spectra that the MLP was trained on. Therefore, some predictions are at the roof or floor of the plot with uncertainty bars cut off, as the MLP has not seen any model spectra either above or below these values.

Jupiter and Saturn spectra were used for two reasons: 1) provide a real application of the MLP's trained on synthetic spectra; and 2) provide an application of testing on spectra outside of one of the parameter's trained grid space (i.e. both Jupiter and Saturn have lower $T_{eff}$ then the models used to train the networks). Users may investigate theoretical/actual exoplanets that lay outside of the trained-on grid space for one or more parameter. Since Jupiter and Saturn are cooler than the models used, the MLP correctly estimated their temperatures to be on the low end of the range. The other three parameters for Jupiter and Saturn were still within the trained-on grid space.

**Figure 7(a):** [M/H]

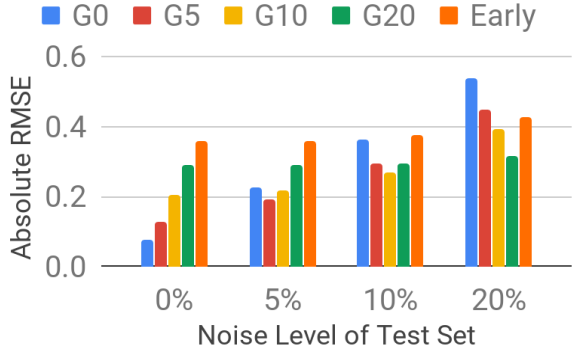

| MLP | 0% | 5% | 10% | 20% |
|---|---|---|---|---|
| G0 | 0.08 | 0.23 | 0.36 | 0.54 |
| G5 | 0.13 | 0.19 | 0.29 | 0.45 |
| G10 | 0.20 | 0.22 | 0.27 | 0.39 |
| G20 | 0.29 | 0.29 | 0.29 | 0.32 |
| Early | 0.36 | 0.36 | 0.38 | 0.43 |

**Figure 7(b):** log $g$ (cm s$^{-2}$)

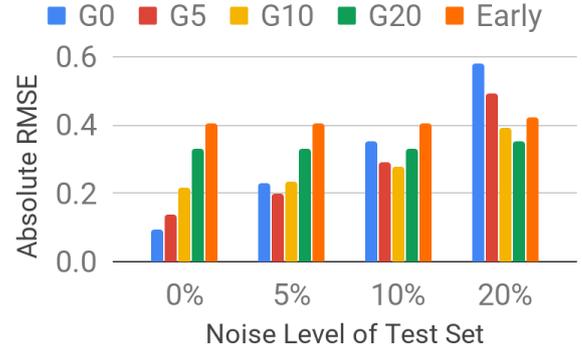

| MLP | 0% | 5% | 10% | 20% |
|---|---|---|---|---|
| G0 | 0.10 | 0.23 | 0.35 | 0.58 |
| G5 | 0.14 | 0.20 | 0.29 | 0.49 |
| G10 | 0.22 | 0.24 | 0.28 | 0.39 |
| G20 | 0.33 | 0.33 | 0.33 | 0.35 |
| Early | 0.41 | 0.41 | 0.41 | 0.42 |

**Figure 7(c):** $T_{\text{eff}}$ [Kelvin]

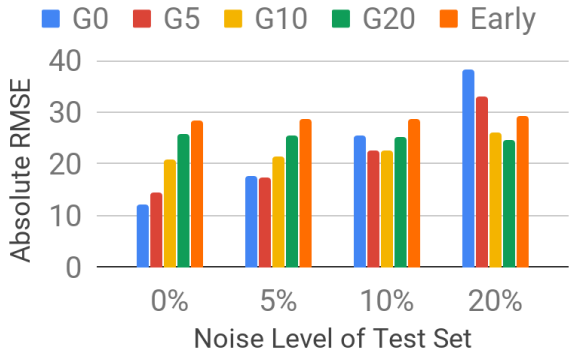

| MLP | 0% | 5% | 10% | 20% |
|---|---|---|---|---|
| G0 | 12 | 18 | 26 | 38 |
| G5 | 14 | 18 | 23 | 33 |
| G10 | 21 | 21 | 23 | 26 |
| G20 | 26 | 26 | 25 | 25 |
| Early | 28 | 29 | 29 | 29 |

**Figure 7(d):** $f_{\text{sed}}$

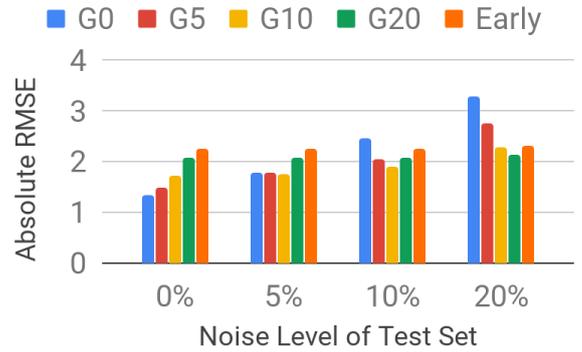

| MLP | 0% | 5% | 10% | 20% |
|---|---|---|---|---|
| G0 | 1.4 | 1.8 | 2.5 | 3.3 |
| G5 | 1.5 | 1.8 | 2.1 | 2.8 |
| G10 | 1.7 | 1.8 | 1.9 | 2.3 |
| G20 | 2.1 | 2.1 | 2.1 | 2.1 |
| Early | 2.3 | 2.3 | 2.3 | 2.3 |

**Figure 7:** Testing results for $\mathbf{S}^{\text{test}}$ after training on $\mathbf{S}^{\text{train}}$ and using $\mathbf{S}^{\text{validation}}$ to stop training with either the greedy or early approach. G0, G5, G10, and G20 refer to greedy stopping optimized to the respective percent of added noise. Testing error varied for each stopping approach by the level of noise in the test set, indicated by the percent seen in the x-axis. For example, in Figure 7(a) the far-left red bar corresponding to G5 is the absolute RMSE in [M/H] measured from testing spectra with 0% noise added to it using the MLP trained greedily for 5% noise. The tables in each panel list the exact testing RMSE values illustrated in the bar plots, which can be used as uncertainty in final output.

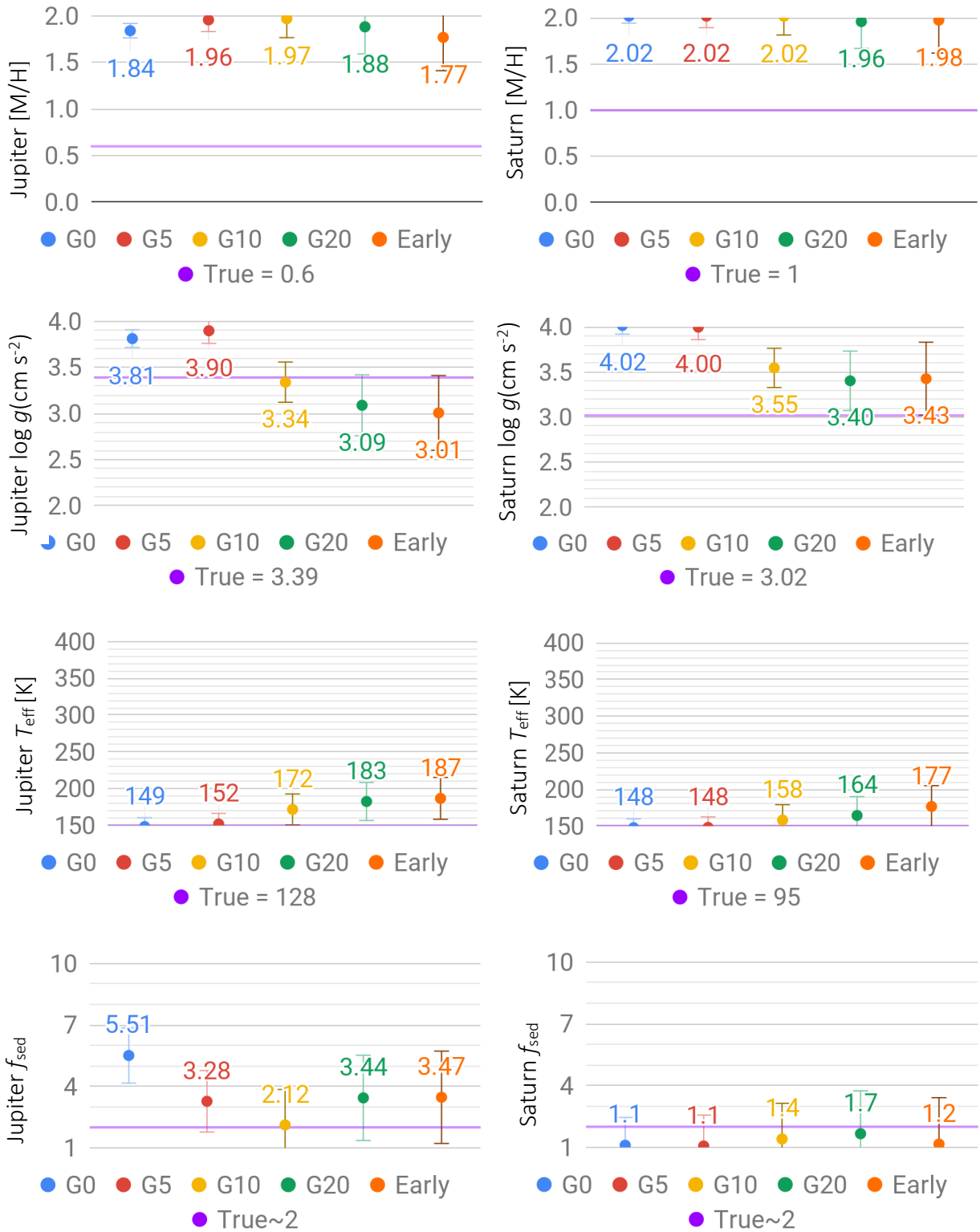

**Figure 8**: Predictions from the five MLP's on the observed Jupiter and Saturn albedo reflectance spectra. The minimum and maximum values on the graphs illustrate the limitations of our predictions, set by the model domain. The horizontal gridlines illustrate the intervals used by the model domain. The solid purple lines are the expected values.

**8: Conclusion**

We found that an MLP is an elegant supervised machine learning approach for rapidly interpreting albedo spectra arising from simulated observations. The five MLPs described here successfully predicted a planet's metallicity, gravity, $T_{eff}$, and cloud properties based on the planet's reflectance spectrum. A tool such as those explored here could be deployed online in concert with observation simulation tools to provide an interpretation of a given simulated spectrum. The MLP thus can provide the user with insight, within less than a second for the cases explored here, into whether a given signal-to-noise ratio is "good enough" to meet the scientific goals desired, for example identifying the correct atmospheric metallicity.

Depending on the simulated noise levels, testing RMSE in these predictions ranged from 0.077 to 0.540 [M/H], 0.096 to 0.580 in log $g$ (cm s$^{-2}$), 12 to 38 in $T_{eff}$ [K], and 1.4 to 3.3 in $f_{sed}$. The RMSE, used as uncertainty in predictions, is on the order of the grid spacing for parameters used to train the networks: $\Delta$ [M/H] = 0.5, $\Delta$ log $g$ = 0.1 cm s$^{-2}$, $\Delta$ $T_{eff}$ = 10 K, and $\Delta$ $f_{sed}$ = 1. When used to predict parameters on Jupiter and Saturn, the MLP's struggled to identify the correct metallicity despite this parameter having the smallest testing error of the four output parameters, relative to the model domain intervals. This is likely a consequence of the real planets lying outside of the model domain and the fact that the models do not yet perfectly reproduce the spectra of these planets. Since the applications we envision for these models, testing the science value of a given simulated spectrum, would use domain models as the input to the simulator, such situations would not arise in practice. Future extensions of this work could consider whether a larger and more robust model domain could robustly interpret the spectra of these planets.

When using one of the five presented MLP's, the user can consider Figure 7 when deciding which MLP to use. The decision should be based on desired accuracy, desired output parameters, and estimated noise levels of the spectra. If the user knows the spectra have high levels of noise, they can get more accurate results by using either Greedy10 or Greedy20. However, if they know the spectra have low levels of noise, it would be better to use either Greedy0 or Greedy5. If the noise level is unknown, we advise to use Greedy20 as it shows the best trade-off, even over the early stopping MLP.

New MLP's can be trained to satisfy different spectra dimensions and desired output parameters, within minutes to hours depending on the size of the MLP. Ultimately, an MLP is a quick and robust means for in-situ estimations. We infer that future work will combine the different machine learning methods circulating in the community, to create an artificial intelligence robust in predicting many different output parameters of planetary spectra at various resolutions, wavelength ranges, and levels of noise; along with combining algorithms to output both a continuous value with uncertainty and posterior distribution for each output parameter.

**9: Acknowledgements**

Support for this research was provided by the Sellers Exoplanet Environment Collaboration. This research would not have been possible if not for the efforts by Virginia C Gulick, from SETI Institute and NASA Ames Research Center, who conducted related research and introduced the idea of this collaboration.

**10: References**